\documentclass[a4paper,pra,twocolumn,showpacs,preprintnumbers]{revtex4}

\usepackage{amsbsy}
\usepackage{amsfonts}
\usepackage{amssymb}
\usepackage{amsmath}
\usepackage{graphicx,color}
\usepackage{graphicx}

\newcommand{\ket}[1]{| #1 \rangle}
\newcommand{\bra}[1]{\langle #1 |}

\begin{document}


\title{Few emitters in a cavity: From cooperative emission to individualization  }

\author{A. Auff\`eves$^{1}$}
\author{D. Gerace$^{2}$}
\author{S. Portolan$^{1}$}
\author{A. Drezet$^{3}$}
\author{M. Fran{\c{c}}a Santos$^{4}$}
\affiliation{$^{1}$CEA/CNRS/UJF Joint Team
``Nanophysics and Semiconductors'', Institut N\'eel-CNRS, BP 166, 25
Rue des Martyrs, 38042 Grenoble Cedex 9, France}
\affiliation{$^{2}$Dipartimento di Fisica
``Alessandro Volta'' and UdR CNISM, Universit\`a di Pavia, via Bassi
6, 27100 Pavia, Italy}
\affiliation{$^{3}$Institut N\'eel-CNRS, BP 166, 25
Rue des Martyrs, 38042 Grenoble Cedex 9, France}
\affiliation{$^{4}$Departamento de F\'isica, Universidade Federal de Minas
Gerais, Belo Horizonte, CP 702, 30123-970, Brazil}

\begin{abstract}
We study the temporal correlations of the field emitted by an
electromagnetic resonator coupled to a mesoscopic number of
two-level emitters that are incoherently pumped by a weak external drive. 
We solve the master equation of the system for increasing number of emitters 
and as a function of the cavity quality factor, and we identify three main regimes 
characterized by well distinguished statistical properties of the emitted radiation. 
For small cavity decay rate, the emission events are uncorrelated and the number 
of photons in the emitted field becomes larger than one, resembling the build-up 
of a laser field inside the cavity. 
At intermediate decay rates (as compared to the emitter-cavity coupling) and for few emitters, 
the statistics of the emitted radiation is bunched and strikingly dependent on the parity 
of the number of emitters. The latter property is related to the cooperativity of the emitters 
mediated by their coupling to the cavity mode, and its connection with steady state subradiance 
is discussed. 
Finally, in the bad cavity regime the typical situation of emission from a collection of individual 
emitters is recovered. We also analyze how the cooperative behavior evolves as a function 
of pure dephasing, which allows to recover the case of a classical source made of an ensemble 
of independent emitters, similar to what is obtained for a very leaky cavity. 
State-of-art techniques of Q-switch of resonant cavities, allied with the recent capability
to tune single emitters in and out of resonance, suggest this system as a versatile source 
of different quantum states of light.
\end{abstract}

\maketitle

\section{Introduction}

The interaction of an ensemble of identical two-level emitters with
the electromagnetic field has been the subject of intense
theoretical and experimental studies since the pioneering work of
Dicke \cite{dicke}. Among the others, superradiance (or
superfluorescence) is the most celebrated effect; it is manifested as a
transient effect by the delayed emission of an intense burst of light,
after the atomic population has been totally inverted \cite{gross_haroche}.
In essence, superradiance is characterized by the emergence of
collective coherence, which leads to a strongly
enhanced emission rate as compared to the case of
uncorrelated emitters. Such striking consequences of the quantum
interference effect in radiation-matter coupled systems has been the subject 
of an intense experimental interest both for atomic \cite{gross_haroche,gibbs77} 
and solid state \cite{scheibner07np} ensembles. 
Early studies on the statistical properties of the emitted light have evidenced 
the role of quantum fluctuations in the initial stages of the superradiant 
emission \cite{bonifacio,Bonifacio_PRA71,clemens02pra}.
Whereas superradiance usually appears in the transient regime,
recent theoretical studies have evidenced other signatures of
collective behavior in the steady state regime, e.g. when  
the two-level emitters are continuously excited by incoherent 
light~\cite{woggonPRL,woggonOE,hollandPRA1,hollandPRA2}.
In these references, cooperative emission is obtained by coupling the
emitters to a low quality ($Q$) cavity mode, which behaves as a 
collective channel of losses.
As a matter of fact, strong confinement enhances the interaction of each
emitter with the cavity field, eventually making it predominant with respect
to individual relaxation channels. Collective dynamics can be
induced this way, provided the atoms identically interact with the
cavity mode, a condition much less stringent than in free space.
Tuning the pumping power allows one to explore different regimes,
leading to different intensities and statistics for the emitted
radiation \cite{hollandPRA1,hollandPRA2}. In particular, it has been shown
that when the pumping strength is too low to saturate a single emitter of the
ensemble, the collection of two-level systems emits a highly 
bunched field~\cite{woggonOE,hollandPRA2}.
The strong temporal correlation has been attributed to the 
relaxation of the system in the subradiant states of the ensemble of emitters. 
These states behave as sources of photon pairs triggered by the driving
pulses, hence the bunching effect in the correlation properties.

In this paper, we analyze the subradiant regime and go 
beyond previous works by keeping the cavity mode as a relevant 
degree of freedom. 
This situation is new with respect to previous theoretical studies, 
where the radiation mode was commonly eliminated by an
adiabatic approximation. We show that the bunching degree decays
with the number of emitters and oscillates with its parity, being
larger when the number of emitters is even, upon converging to a
single value as the number of emitters grows. We give an
interpretation of this effect, based on the structure of Hilbert space 
of the system. 
Furthermore, keeping the cavity mode in the description allows
to explore two other limits, depending on the cavity finesse.
If this parameter is significantly increased, the strong coupling
conditions lead to the emission of a field showing no temporal
correlations, an effect that we identify with the emission of
Poissonian radiation similar to a lasing transition. Recently, such
regime has been investigated from the spectral point of view
\cite{poddubny2010,laussy2011preprint}, while here we address its 
statistical properties. On the contrary, for decreasing cavity finesse one
progressively decreases the coupling with the mode, allowing to
continuously explore the boundary between collective and individual
behavior of such radiation sources. 
When all the emitters become uncorrelated, increasing their number 
one by one up to large values coincides essentially with building 
a classical source with chaotic statistical properties. 
Dipole correlation can also be destroyed with pure dephasing, 
a situation that we eventually explore as well.

\begin{figure}[t]
\begin{center}
\includegraphics[width=0.48\textwidth]{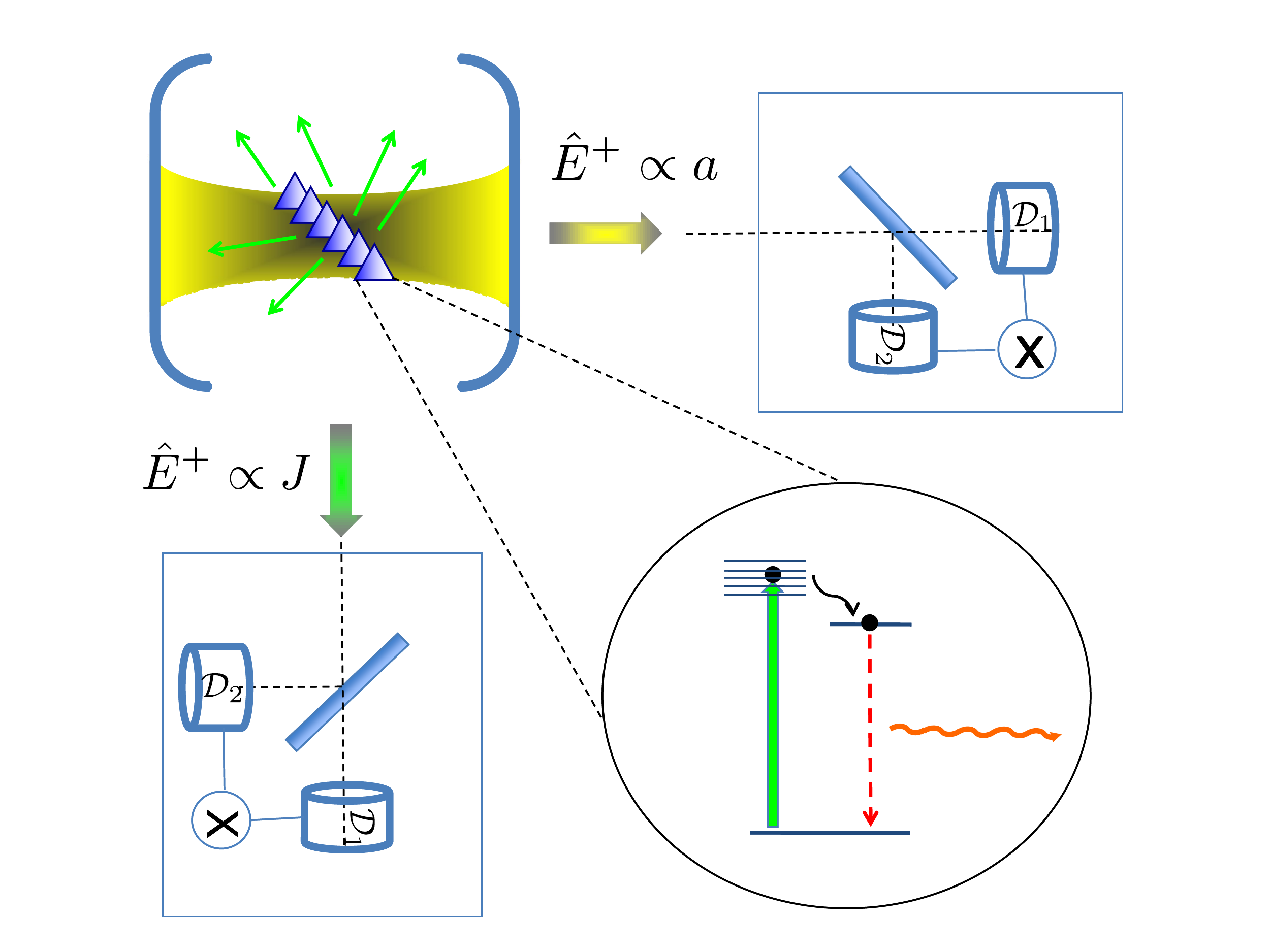}
\caption{Schematic of the system under consideration in this work:
$N$ two-level emitters incoherently pumped by an external drive,
coupled to a single resonator mode. Correlations in the emitted
field can be detected through a HBT set-up either from the cavity
or from the emitters' loss channel, respectively. A typical
incoherent pumping scheme is shown, in which high-energy excitations
relax down to a well defined ground state transition (e.g. through
phonon scattering) via photon emission. }\label{fig1}
\end{center}
\end{figure}

The present work becomes particularly relevant when considering
that the subradiant regime is reachable if and only if each single
emitter of the ensemble is predominantly coupled to the cavity mode.
In particular, this allows to pump each emitter very slowly, but
still faster than any individual relaxation mechanism. This is nowadays
possible thanks to impressive progress in nano-fabrication
technology on single emitter-cavity 
coupling~\cite{strauf06,dicarlo2010nat,wallraff04,reithmaier04,yoshie04,peter05,lethomas06,
kaniber08,park06,hennessy07,faraon08,suffczynski09,munsch09,jacek10}.
Small assemblies of artificial quantum emitters, like quantum
dots \cite{strauf06} and superconducting qubits
\cite{dicarlo2010nat} coupled to a high-Q resonator mode have
recently become available, strongly motivating to investigate the
new physics provided by these systems. Moreover, beyond introducing
an external parameter to control the steady state collective
regime, this model is of great interest also for the study of lasing
transition, as well as to investigate the connections between
superradiance and lasing \cite{haake93prl}. 

The paper is organized as follows. We first introduce the model and 
the methods employed to solve it, both numerically and analytically 
(Sec.~II). Then, we present results as a function of cavity finesse, 
and we give a physical interpretation for the different regimes found
(Secs.~III, IV, and V). 
Before summarizing the main conclusions, we give a brief description of
possible realistic implementations to explore the physics
investigated in this work (Sec.~VI).

\section{Model and methods}

The system under study consists of $N$ two-level emitters coupled to
the mode of an electromagnetic cavity of annihilation (creation)
operator $a$ ($a^\dagger$) with equal coupling strength, $g$, and it
is schematically pictured in Fig.~\ref{fig1}. To focus on collective
effects, we will assume that the resonator and all the emitters have
the same transition frequency, $\omega$. The range of validity of
this ideal model in practical implementations is discussed in Sec.
VI. The model can be rigorously described by the Tavis-Cummings
Hamiltonian (in the rotating wave approximation)
\begin{equation}\label{TC_H}
H=\sum_{i=1}^N \hbar \omega (\sigma_{z}^i) + \hbar \omega a^\dagger
a + i g (a J_+ - J_- a^\dagger)\, ,
\end{equation}
where $J_+=\sum_i \sigma_+ ^i$ is the collective dipole operator,
and $\sigma_- ^i=|g_i\rangle\langle e_i|$ is the lowering operator of
the single emitter $i$, whose excited (ground) state is denoted by
$|e_i\rangle$ ($|g_i\rangle$). The Hilbert space corresponding to
the assembly of resonant atoms is equivalent to the tensor product
of the 2-dimensional space of each single emitter, i.e. $\mathcal{H}
\simeq (\mathbb{C}^2)^{\otimes N}$, spanned by the basis of
factorized states in the form $| i_1 \rangle | i_2 \rangle ... | i_n
\rangle$, where $i_n = \{g_n,e_n\}$ (in the following we shall refer to
this basis as the uncoupled basis).
When including possible dissipation channels and external driving,
the Markovian dynamics of the full system is given by
\begin{equation}\label{ME}
\dot{\rho}=-i[H,\rho] + \sum_{i=1}^N \mathcal{L}_i(\rho) +
\mathcal{L}_a (\rho)+  \mathcal{L}_{\gamma^*}(\rho) \, .
\end{equation}
The Lindblad dissipative terms, modeling incoherent loss and pumping
of each emitter, explicitly read
\begin{eqnarray}\label{Pgamma}
\mathcal{L}_i(\rho) &=& \mathcal{L}_{\gamma} (\rho) + \mathcal{L}_{P_x} (\rho) \nonumber \\
&=& \frac{\gamma}{2}(2\sigma_-^i \rho
\sigma_+^{i} - \sigma_+^{i} \sigma_-^i \rho - \rho \sigma_+^{i} \sigma_-^i) + \nonumber \\
&& \frac{P_x}{2}(2\sigma_+^{i} \rho \sigma_-^i - \sigma_-^i \sigma_+^{i} \rho
- \rho \sigma_-^i \sigma_+^{i})\, ,
\end{eqnarray}
where the quantity $\gamma$ is the individual emission rate in loss
channels distinct from the cavity mode for each emitter, and $P_x$
quantifies the excitation rate. As pictured in Fig.~\ref{fig1},
incoherent pumping can be obtained by resonantly exciting higher
energy states, that further relax down to the levels of interest.
The decay of the cavity mode with rate $\kappa$ through imperfect
mirrors is described by the following term
\begin{equation}\label{a}
\mathcal{L}_a(\rho) = \frac{\kappa}{2}(2 a \rho
a^\dagger - a^\dagger a \rho - \rho a^\dagger a)\,
\end{equation}
whereas the pure dephasing of the emitters, which is unavoidable in
solid-state artificial atoms, can be taken into account by a
Liouvillian term of the form \cite{pra}
\begin{equation}\label{gammastar}
\mathcal{L}_{\gamma^*}(\rho) = \sum_{i=1}^N
\frac{\gamma^{\ast}}{2}(\sigma_{z}^i \rho \sigma_{z}^i - \rho ) \, .
\end{equation}
Following standard input-output theory \cite{Gardiner_Zoller}, the
output field at the photodetectors is proportional to the $a$
operator (and its conjugate). We are interested in the temporal
correlations of the field radiated by the cavity mode.
Experimentally, an ordinary Hanbury-Brown Twiss (HBT) setup can be
used, giving access to the steady state second-order correlation at
zero-time delay \cite{paper_g2}
\begin{equation}\label{g2_a}
g_{a}^{(2)}(0)=g_{ss}^{(2)}=\langle a^\dagger a^\dagger a a\rangle /
\langle a^\dagger  a \rangle^2 \, .
\end{equation}
In order to calculate the latter quantity, we numerically solve 
Eq.~(\ref{ME}) either by direct integration or by Quantum Monte
Carlo (QMC) simulations \cite{dalibard,carmichael_book}. Direct
integration allows one to obtain accurate results at lower computational
time costs, for which we use an extension of the code already employed
in a previous work~\cite{ours}. 
However, the tensor structure of the phase space makes the
computational resources needed for a direct integration rapidly
diverge. On the contrary (wave function) QMC gives the solution as a
stochastic average over quantum trajectories of wavefunctions
{unravelings}, drastically reducing the amount of space needed and
allowing us to push forward the previous computational limits
\cite{carmichael_book}.

If $\kappa$ is large enough, the cavity mode can be adiabatically
eliminated from the equations, giving rise to an effective dynamics
described by the reduced master equation
\begin{equation}\label{ME_adiab}
\dot{\rho}=\mathcal{L}_{\Gamma} (\rho) + \sum_{i=1}^N
\mathcal{L}_i(\rho) \, ,
\end{equation}
where
\begin{equation}\label{J}
\mathcal{L}_{\Gamma} (\rho) = \frac{\Gamma}{2}(2J_- \rho
J_+- J_+ J_- \rho - \rho J_+ J_-)\, .
\end{equation}
The physical meaning of this equation is the following: the cavity
acts as a common relaxation channel for the atoms, whose dissipation
rate is quantified by the coupling strength with the collective
atomic dipole $J_-$, i.e. $\Gamma = 4g^2/\kappa$. Namely, adiabatic
elimination is valid when $\kappa \gg N\Gamma$ \cite{hollandPRA1}.
In this regime, the cavity field, and thus the output field is
proportional to the collective dipole degree of freedom, $J_-$.
Consequently, the temporal autocorrelation function at zero time
delay can be expressed as
\begin{equation}\label{g2_J}
g_{J}^{(2)}(0)=\langle J_+ J_+ J_- J_-\rangle / \langle J_+ J_-
\rangle^2 \, .
\end{equation}
Note that the same detection properties could be obtained by
registering the far field where the atomic dipoles interfere, as it
is pictured in Fig.~\ref{fig1}. 
Also for Eq. (\ref{g2_J}) we can employ numerical approaches either based
on direct integration or on QMC, respectively. 
As usual, once the master equation is solved
for $\rho$ the expectation values can be calculated
with the general rule $\langle \hat{O} \rangle=\mathrm{Tr}\{\hat{O}\rho\}$, 
where $\hat{O}$ identifies the operator corresponding to a given 
observable.

Sweeping the cavity finesse allows to continuously
explore the transition from a collective behavior driven by
$\mathcal{L}_{\Gamma} (\rho)$ to an individual one, corresponding to
$\mathcal{L}_i(\rho)$. In this work we analytically study some
limits of these regimes. If the dynamics is individual, it is naturally 
described in the uncoupled basis mentioned above. If the dynamics 
is collective, an order appears in the system due to the build up of 
correlations between dipoles belonging to different atoms \cite{gross_haroche},
manifested by the occurrence of coherences in the uncoupled basis.
In the latter case, a natural basis is provided by the eigenstates
$\ket{J,M}$ of the total angular momentum $\{J^2, J_z\}$ (in the
following, we shall talk about the coupled basis of the atomic
states). Note that among them, the states satisfying $J_-\ket{J,M}=0$ 
are not coupled to the electromagnetic field and are thus defined as
``subradiant''.

\section{Second-order correlations of the emitted field}

\begin{figure}[t]
\begin{center}
\includegraphics[width=0.48\textwidth]{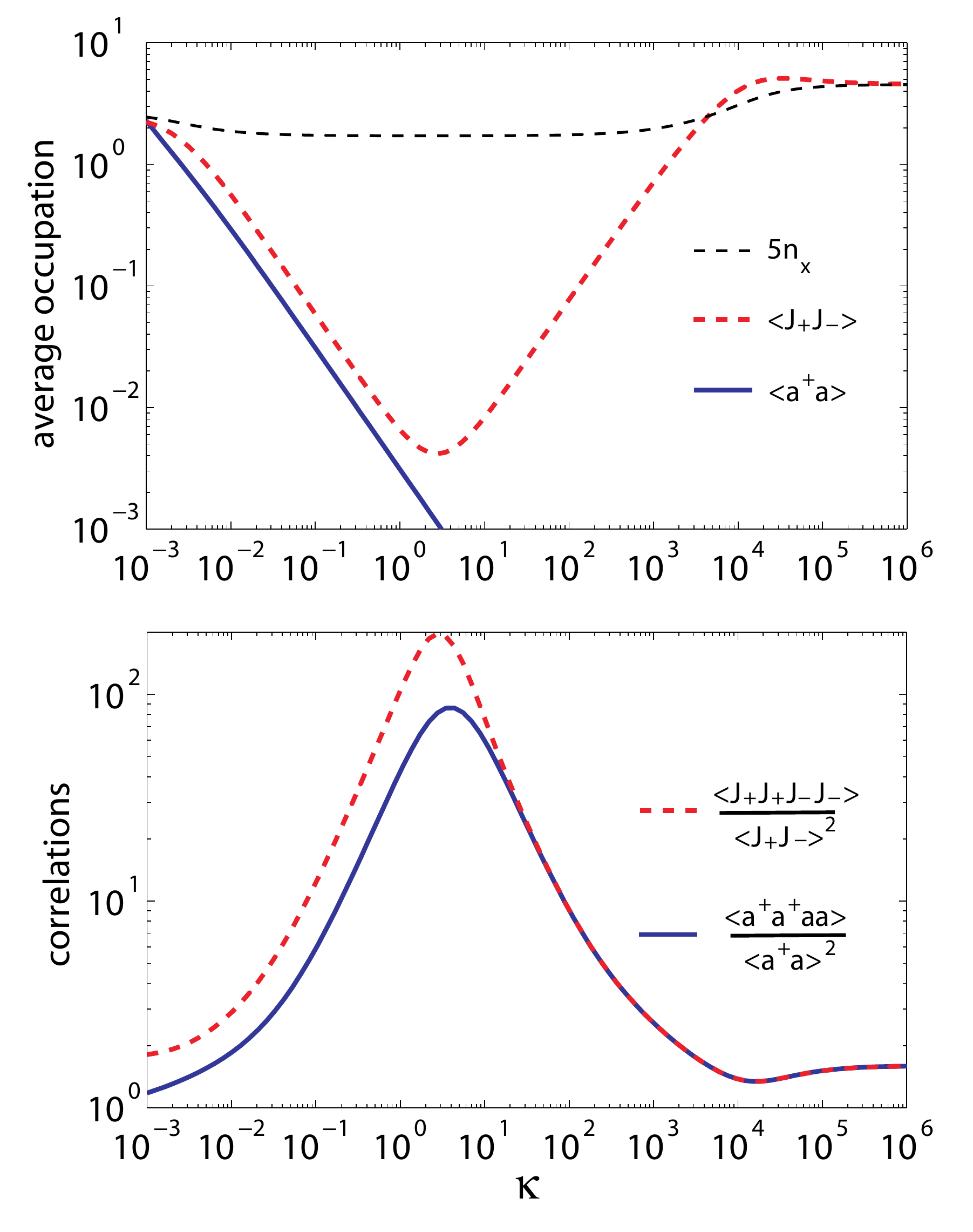}
\caption{Full numerical results for average populations and
second-order correlation in steady state as a function of $\kappa$
(normalized to $g$) for the case of $N=5$ atoms in the cavity,
calculated from atomic and cavity fields, respectively. Parameters
are: $P_x=0.001g$, $\gamma=0.0001g$. } \label{fig2}
\end{center}
\end{figure}

As mentioned above, the regime studied in the literature so far is the
adiabatic elimination regime, where the cavity mode operator can
be assumed as $a \propto J_-$. 
We assume a fixed and weak incoherent pumping rate, $P_x /g$.
In order to understand the collective behavior and to link it to the previous literature, 
we plot in Fig.~\ref{fig2}a the average cavity mode occupation, 
$n_a=\langle a^\dagger a \rangle$, the average collective occupation of the 
emitters, $n_J=\langle J_+ J_- \rangle$, and the total emitters population, 
$N n_x $ (where  $n_x=\langle \sigma_{+}^{i} \sigma_{-}^{i} \rangle$ 
in the steady state.
The second-order correlations are plotted in Fig.~\ref{fig2}b, and they 
have been calculated both from the cavity mode operator, Eq. (\ref{g2_a}), 
and the collective dipole operator, Eq. (\ref{g2_J}), respectively.   
The results are shown for $N=5$ emitters in the cavity, which 
represents a good compromise to simultaneously have a well converged 
numerical solution for Eq. (\ref{g2_a}) and to evidence the effects of collective 
behavior. 
As it can be seen in Fig. ~\ref{fig2}b, the two calculated correlations don't 
match for most of the parameter space, showing that the cavity mode has 
to be taken into account to obtain the correct result when the cavity finesse 
is large. The two curves converge to the same value when $\kappa/g \simeq 50$, 
where the adiabatic approximation starts to be satisfied.

We first focus on the case where adiabatic elimination is valid: two
different situations emerge. The case where $\Gamma = 4g^2 /\kappa >
\gamma$ corresponds to a collective dynamical behavior for the atomic
ensemble. In the low pumping regime under study, the emitted field
is strongly bunched, $g^{(2)}(0) > 1$, confirming previous
theoretical works  \cite{woggonOE,hollandPRA2} on a similar model. 
In the two-atom case, this behavior has been
attributed to the emission of so called ``super-radiant'' photon pairs
\cite{woggonOE}, typically separated by a delay scaling like the
inverse of the pumping rate. This regime has been further explored
for a larger number of emitters in \cite{hollandPRA1,hollandPRA2}, 
where the authors show that the atomic ensemble has mostly relaxed 
in the subradiant states. This is in agreement with Fig.~\ref{fig2}a where
the average intensity emitted by uncorrelated atoms, $N n_x$ is 
compared to the average fluorescence field emitted by the collective
atomic dipole, $\langle J_+ J_- \rangle$. 
It can be clearly observed that in a $\kappa /g$ range spanning
several decades, the condition $\langle J_+J_- \rangle < N
n_x$ is usually satisfied, a signature of the anticorrelation of the atomic 
dipoles and of the efficient relaxation in the sub-radiant states.
Note that the subradiant regime is reached when $P_x<\Gamma$, a
condition that is not fulfilled anymore for $\kappa/g > 10^3$. In that
case the steady state emission becomes super-radiant, i.e.   
$\langle J_+J_- \rangle > N n_x$, and correspondingly  $g^{(2)}(0)$ 
approaches $1$, a limit that is exactly reached only for $N\gg 1$ as 
evidenced in previous work~\cite{hollandPRA2}. Finally, when 
$\kappa / g$ is increased up to a point where $\gamma > \Gamma$ 
(very bad cavity regime), emission appears as if it was coming from
an independent collection of emitters. Under such conditions, the
limiting values are such that $\langle J_+J_- \rangle \to N n_x$
and $g^{(2)}(0) \to 2(1-1/N)$, a limit that we analyze in detail below.

\begin{figure}[t]
\begin{center}
\includegraphics[width=0.49\textwidth]{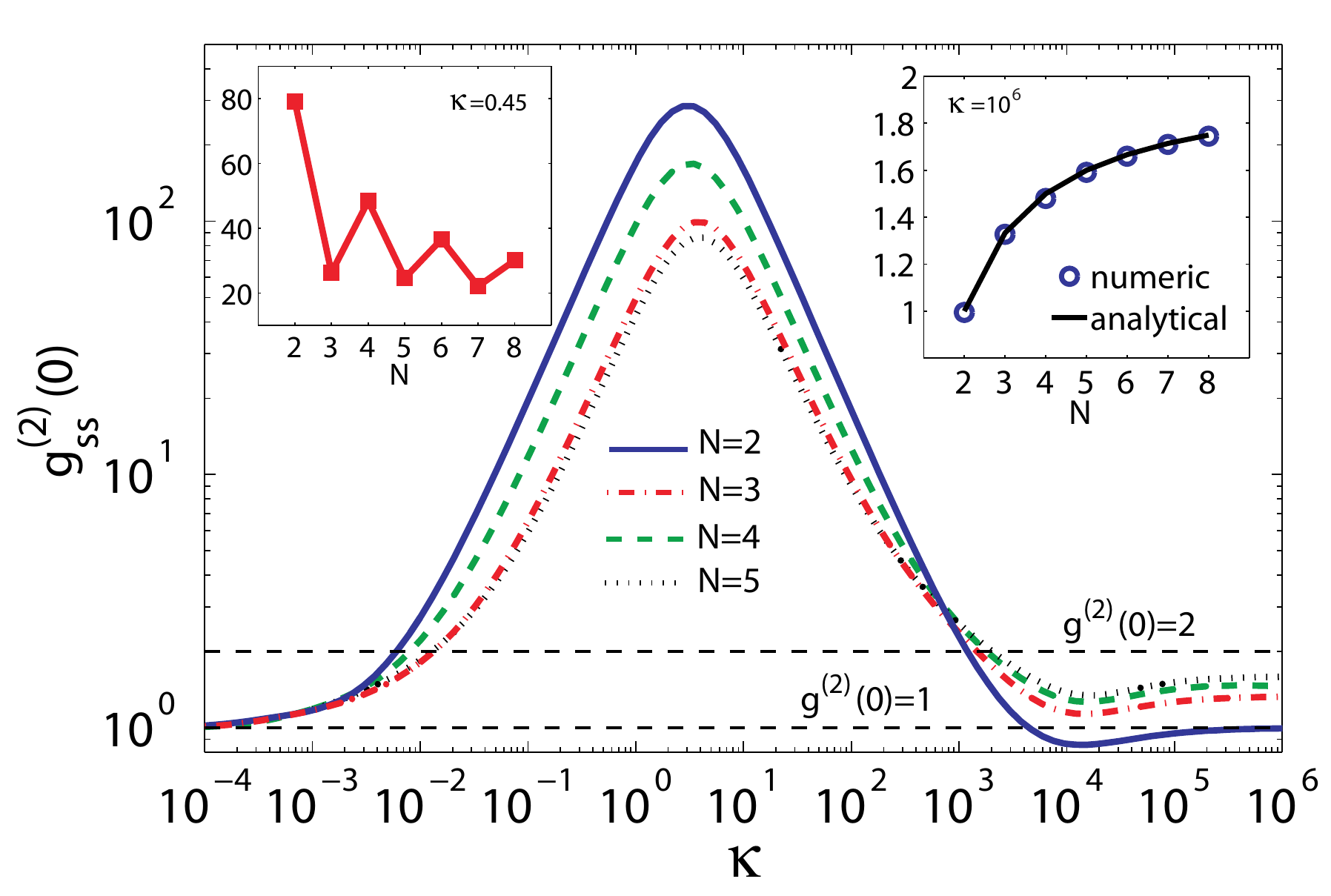}
\caption{ Full numerical solution for zero time delay second-order
correlation in steady state calculated from the cavity degree of
freedom, as a function of $\kappa$ (normalized to $g$), for low
incoherent pump rate $P_x=0.001g$ and emitters' dissipation
$\gamma=0.0001g$, and varying number of emitters, $N$. Left inset:
oscillations as a function of $N$ in the good cavity, subradiant regime. 
Right inset: bad cavity limit of correlations as a function of $N$.}
\label{fig3}
\end{center}
\end{figure}


We focus now on the statistical properties of the system by looking at the 
dependence on the number of emitters, the results are shown in Fig.~\ref{fig3}
with the same parameters used in Fig.~\ref{fig2}.
Keeping the cavity mode as a relevant degree of freedom allows one to 
explore the temporal correlations on the entire $\kappa/g$ range, which 
can only be done by the full numerical solution of the master equation. 
At this point, we go beyond previous works. 
First, we notice that a transition to amplified spontaneous emission clearly
occurs for very large cavity finesse, i.e. as soon as   $\kappa/g < 0.001$ 
($P_x>\kappa$), with  the temporal fluctuations becoming Poissonian 
independently of the number of emitters.  A temporal picture for 
this effect is that as soon as $P_x>\kappa$, the cavity stores the 
photons on a timescale longer that the typical delay between two 
emission events by the atomic assembly, blurring out any temporal 
correlation feature. This situation has been already evidenced for the 
single emitter case in \cite{ours,Delvalle,Elena}, and it was identified 
as a lasing threshold signature. 
Here, the system becomes a source of radiation with Poissonian 
statistical noise, reminiscent of a few emitters laser. The case for $N=2$ in the 
strong emitter-cavity coupling regime was also studied for a similar model 
in Ref.~\cite{yeoman,Elena2007} as a function of pump power. 
As a matter of fact, three main effects on measurable quantities can be identified: 
(i) the average cavity population becomes larger than one, setting the 
onset of stimulated emission (see also Fig.~\ref{fig2}a), 
(ii) the population of each emitter is clamped to a definite value, here, 
$n_x \simeq 0.5$  (i.e. $5n_x \to 2.5$ in Fig.~\ref{fig2}a), and (iii) 
$g^{(2)}(0) \to 1$ (more clearly evidenced in Fig.~\ref{fig3} for any $N=2,3,4,5$). 

On increasing $\kappa/g$, it is evident from Fig.~\ref{fig3} that 
in the subradiant range there is an unexpected and quite pronounced 
parity-dependent oscillation of the photon correlation value at zero-time 
delay, a behavior that has not been discussed in the literature, to our 
knowledge. 
Such parity oscillations are not limited to the adiabatic 
elimination region (see the inset on the left panel for $\kappa/g =0.45$ 
and up to a number of emitters $N=8$).
The latter calculation required consistent numerical optimization 
and was performed on a cluster, where the full master Eq. (\ref{ME}) 
could be solved without truncation effects on the convergence.
Although quite difficult to be analyzed in the dressed state picture, 
the oscillatory behavior of correlations as a function of the number 
of emitters can be roughly interpreted  as a consequence 
of the collective coupling in Eq.~(\ref{J}), which dominates the dynamics 
over the individual relaxation channels in Eqs.~(\ref{Pgamma}) 
and (\ref{a}). The former couples the states with the same quantum 
number $J$ and is diagonal in the coupled basis. On the contrary, 
the other channels act on each emitter and are responsible for 
coherences between states with different values of $J$.
We will investigate this regime in detail in Sec.~IV. 
The bunching behavior reaches a maximum for $\kappa\simeq g$, slightly
depending on the number of emitters. An interesting even/odd behavior 
has also been found in the spectral response~\cite{poddubny2010} and it
would be interesting to relate the two phenomena in future developments
of the present work.

Finally, in the opposite limit of $\kappa /g$ becoming large enough that
$\gamma > \Gamma$ the radiation is characterized by a field emitted from
independent atoms. Oscillations tend to disappear, as it can be seen in
Fig.~\ref{fig3} (see also inset on the right hand side). 
In this regime, the single emitters undergo individual dynamics,
and the dipoles become uncorrelated (see also  the large $\kappa$
limit in Fig.~\ref{fig2}a, where $\langle J_+J_- \rangle \to N n_x$. 
Restricting our considerations to the correlation function, 
an analytical expression of $g^2(0)$ is derived in Sec.~V, 
where we show that $g^{(2)}(0)=2(1-1/N)$. Note that in the 
limit of a large number of atoms, $g^{(2)}(0)\to 2$, which corresponds 
to the emission of an incoherent chaotic field.

\section{Parity-related oscillations}

In this Section we calculate analytical expressions for the bunching
degree of the emitted field, in the two and three emitters cases,
which provide a physical explanation for the observed parity dependence.
The analytic treatment is valid in the regime of adiabatic elimination of
the cavity mode dynamics. To sustain our analysis, we have also used
a Quantum Monte Carlo (QMC) simulation \cite{carmichael_book}. We
have calculated $g^{(2)}(0)$ as a function of the number of emitters
$N$, with $N$ being as large as $10$. The results are plotted in
Fig.~\ref{fig4}a by simulating the reduced master equation, 
Eq. (\ref{ME_adiab}). As it can be seen, we alternatively recovered 
the result shown in Fig.~\ref{fig3} at a specific value of $\kappa /g =100$, by 
generalizing to a larger number of emitters that would be unaccessible through 
direct numerical integration. Again, bunching and oscillatory behaviors are 
clear signatures of this collective low pumping regime, which we are going
to analytically describe below.

\subsection{Two emitters case}

In the following of this Section we use the coupled basis for the emitters. 
Such basis of states is schematically indicated in Fig.~\ref{fig4}b.
In particular, the two-emitter case is pictured in the first panel of 
Fig.~\ref{fig4}b, and we will focus on this diagram here. 
In this specific case, the basis $\ket{J,M}$ is non-degenerate 
and the Hilbert space consists of the three symmetric (or 
Dicke) states $\ket{1,1}=\ket{e_1,e_2}$,
$\ket{1,0}=(\ket{e_1,g_2}+\ket{g_1,e_2})/\sqrt{2}$,
$\ket{1,-1}=\ket{g_1,g_2}$, and the antisymmetric state
$\ket{0,0}=(\ket{e_1,g_2}-\ket{g_1,e_2})/\sqrt{2}$ (i.e., the usual triplet and 
singlet states of a spin 1 system). 
The steady state (and $g^{(2)}(0)$ statistics) is the result of a competition between
collective (specifically $\mathcal{L}_{\Gamma}$) and individual
decoherence channels
(i.e., $\mathcal{L}_{P_x},\mathcal{L}_{\gamma},\mathcal{L}_{\gamma^*}$).
The results shown in Fig.~\ref{fig4}a are obtained without pure
dephasing and with $P_x= 10 \gamma$. In the steady state, low
pumping equally populates the subradiant states $\ket{1,-1}$ and
$\ket{0,0}$. As it was first discussed by Temnov and Woggon \cite{woggonOE}, 
one can decompose the problem in two mutually
exclusive cycles. In the regime under study, the $\Gamma$-related
de-excitation events are much more probable than the excitation events driven by 
the pumping rate $P_x$. Hence,
as soon as an excitation is stored in a state where the two
mechanisms are possible, it is more likely that the excitation will undergo a
collective emission than an upward pump jump.
The microscopic scenario will then involve one cycle (pictured in
green) consisting of the excitation and relaxation of the symmetric
state $\ket{1,0}$, leading to the emission of single photons. In the
other cycle (pictured in red), the excitation is first stored in
$\ket{0,0}$. Then, during the time evolution, the only possibility is to
excite the state $\ket{1,1}$, leading to the fast emission of two
photons. The system prepared in the subradiant state $\ket{0,0}$
behaves thus like a deterministic source of photon pairs. Each pair
is emitted on a timescale $1/\Gamma$, the delay between two pairs
being typically $1/P_x$. Consequently, the emitted field is expected
to show bunching, $g^{(2)}(0)\sim \Gamma/P_x$, an order of magnitude
that we exactly recover below by analytically solving the approximate
master equation.

In fact, to go beyond this qualitative interpretation we have computed the
expression for $g^{(2)}(0)$ in terms of the relevant elements in the steady
state density matrix
\begin{equation}\label{g2_0N2}
g^{(2)}(0) =  \frac{\rho_{1,1}}{[\rho_{1,1}+\rho_{1,0}]^2}\, ,
\end{equation}
at first order in $P_x/\Gamma$. The diagonal matrix elements of the
density matrix are written with the row (column) index only, e.g.
$\bra{J,M} \rho \ket{J,M} = \rho_{J,M}$. In this two-emitter case
the evolution equations for the diagonal elements do not involve
off-diagonal terms and the solution is obtained by solving a system
of rate equations governing the evolution of the atomic populations.
With lengthy but straightforward calculations, we express Eq. (\ref{ME_adiab}) 
on the basis above, and solve it for  the steady state $\dot{\rho}=0$.
We finally obtain an analytical estimation 
\begin{equation}
g^{(2)}(0) = \frac{16}{9} \frac{g^2}{\kappa P_x} = \frac{4}{9} \frac{\Gamma}{\kappa P_x} \, , 
\end{equation}
which for $P_x /g=10^{-3}$ and $\kappa/g=100$ gives $g^{(2)}(0)\simeq
17.8$, a value that is very close to the numerical results (obtained from both
direct numerical integration and QMC).

\subsection{Three emitters case}\label{3emitters}

\begin{figure}[t]
\begin{center}
\includegraphics[width=0.48\textwidth]{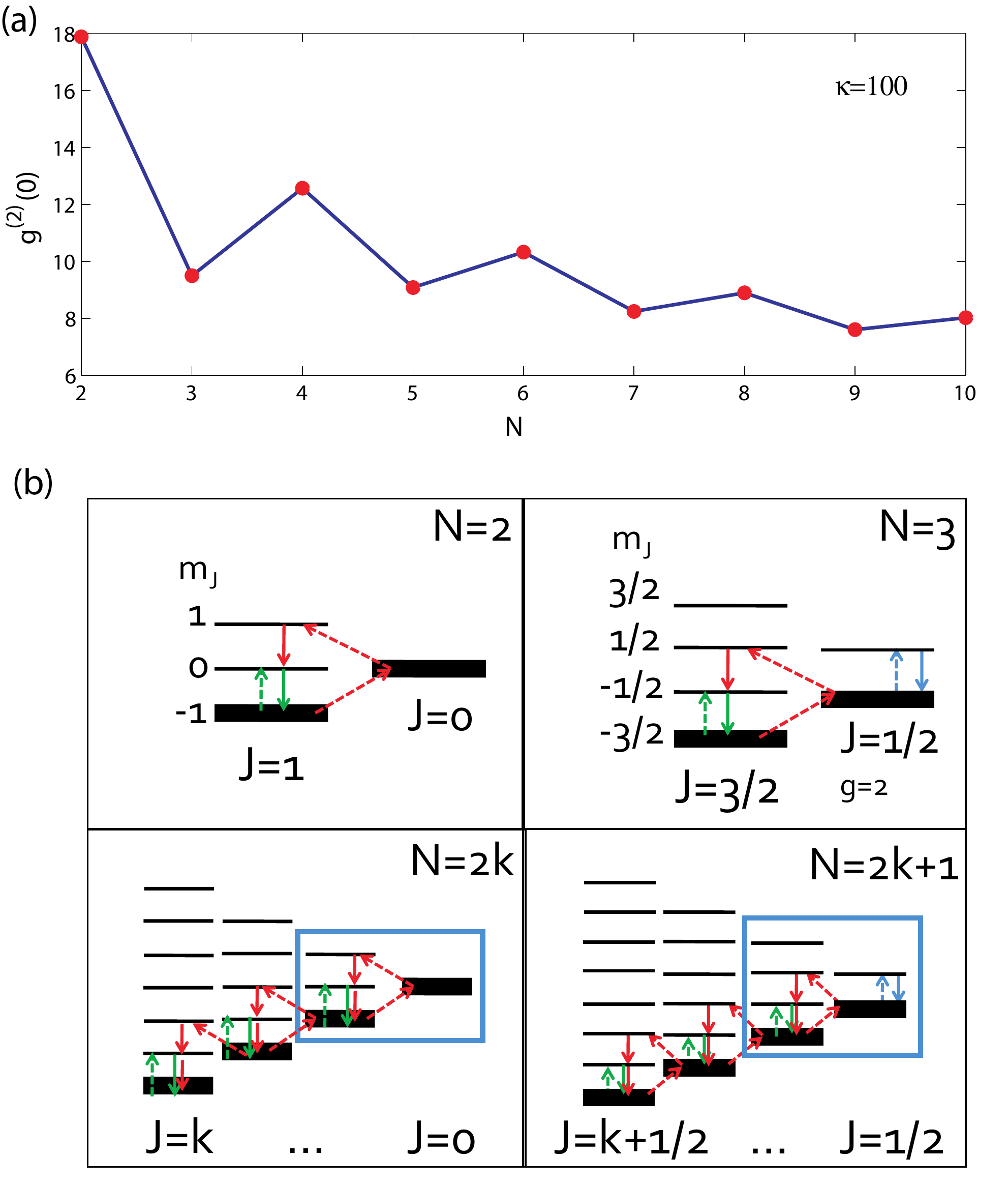}
\caption{(a) Oscillations in the second-order correlation at zero
time delay in the bad cavity, sub-radiant regime ($\kappa/g =100$). 
(b) Schematic
energy level diagram for $N=2$ and $N=3$ in the total angular
momentum basis, and their generalizations for even and odd number of
emitters, respectively. For $N=3$, $g=2$ reminds the two-fold
permutation degeneracy }\label{fig4}
\end{center}
\end{figure}

In the $N=3$ case, the simple picture above cannot be applied
directly. The most important difference emerges immediately from the
very nature of the Hilbert space. As pictured in Fig.~\ref{fig4}b, right panel
above, the $J=J_{\text{min}} = 1/2$ subspaces have a 2-level system
structure. As a consequence, the subradiant states corresponding to
$\ket{J_{min},-M_{J_{min}}}$ do not behave as deterministic sources 
of photon pairs as in the $N=2$ case, but eventually give rise to
single-photon cycles when they are pumped (light blue arrows in
Fig.~\ref{fig4}b). The presence of these extra loops spoils the
bunching of the emitted field, and this gives a simple explanation
for the abrupt loss of $g^{(2)}(0)$ bunching when going from two to three 
emitters. 

To sustain this argument we have computed the second-order correlation 
$g^{(2)}(0)$ in terms of the density matrix elements, which for the $N=3$ case reads
\begin{equation}\label{g2_0N3}
g^{(2)}(0) =  \frac{12 ( \rho_{\frac{3}{2},\frac{3}{2}} +
\rho_{\frac{3}{2},\frac{1}{2}})}{[ 3 \rho_{\frac{3}{2},\frac{3}{2}}
+ 4 \rho_{\frac{3}{2},\frac{1}{2}} + 3 \rho_{\frac{3}{2},-\frac{1}{2}}
+ \rho_{\frac{1}{2},\frac{1}{2};1} + \rho_{\frac{1}{2},\frac{1}{2};2}]^2}\, .
\end{equation}
We followed the same notation as in Eq.~(\ref{g2_0N2}) for the
matrix elements. The subradiant states are written
$\ket{1/2,-1/2;1}$ and $\ket{1/2,-1/2;2}$ \cite{Arecchi_PRA72}, where the second label is
a shorthand notation related to the $g=2$ extra degeneracy (see App.~\ref{A}
and the related figure). After writing the master equation in this basis,
we then proceeded with a perturbative solution of the linear system 
for the steady-state matrix elements in powers of $(P_x/\Gamma)$.
Differently from the $N=2$ case, the equations of motion for the diagonal
terms now involve off-diagonal matrix elements, preventing one from
describing the dynamics with a simple set of rate equations.
Nevertheless, within the same perturbative picture, we can consider
these terms to be much smaller than the diagonal ones. Indeed, the
coherences are brought by some residual individualization of the
$P_x$-driven channel that, for the parameters chosen, acts as a
perturbation over the collective dynamics induced by the $\Gamma$
events.
The latter argument has been checked with the steady-state density
matrix elements obtained numerically.
As detailed in App.~\ref{analytics}, the final result reads
\begin{eqnarray}\label{g2N3t}
g^{(2)}(0) \simeq 
&&  \frac{14}{75} \frac{\Gamma}{P_x} + 2.067 - 3.501 \frac{P_x}{\Gamma} 
+ 9.18 \left( \frac{P_x}{\Gamma} \right)^2 \nonumber \\
&& - 23.4 \left( \frac{P_x}{\Gamma} \right)^3 + \mathcal{O}\left( \frac{P_x}{\Gamma}\right)^4\, .
\end{eqnarray}
We obtain the analytical value $g^{(2)}(0) \simeq 9.46$, which
compares very well with the numerically calculated value $\simeq 9.55$ 
reported in Fig.~\ref{fig4}a  from QMC simulations. 
Notice that the population of the state $\ket{\frac{3}{2},\frac{3}{2}}$ 
cannot be neglected and was also taken into account in the 
analytical calculation.

\subsection{Generalization to $N>3$ emitters}

Considering the energy level diagrams in the generic even and odd
cases, as depicted in Fig.~\ref{fig4}b, lower panels, one can always 
find the two patterns above (i.e., $N=2$ and $N=3$) as distinctive features 
of each scenario.
For increasing number of emitters more subradiant states
become available, among which the population arising from the $P_x$
reservoir will be shared (bold levels in the figure). This spreading of
population unavoidably leads to the decay of the bunching degree 
as $N$ is increased, as observed in Figs.~\ref{fig3} and \ref{fig4}a, 
and confirmed by the general trend anticipated in \cite{woggonOE} 
where a semi-classical approach was used. 
In the limit of large large numbers, such as $N=10000$, the bunching 
degree of $g^{(2)}(0)$ converged to $2$, as for classical chaotic sources.
Note that in our case taking the limit of large numbers requires
some caution, as one could exit the validity of the adiabatic elimination 
approximation, $\kappa \gg N\Gamma$, and eventually enter strong collective
light-matter coupling regime (see, e.g., recent works \cite{SCNemitters,Aigor}). 
For the parameters chosen in the simulation in Fig.~\ref{fig4}a, the condition for
adiabatic elimination limits the maximum number of emitters 
to about 2000.

\section{Individualization}

\begin{figure}[t]
\begin{center}
\includegraphics[width=0.5\textwidth]{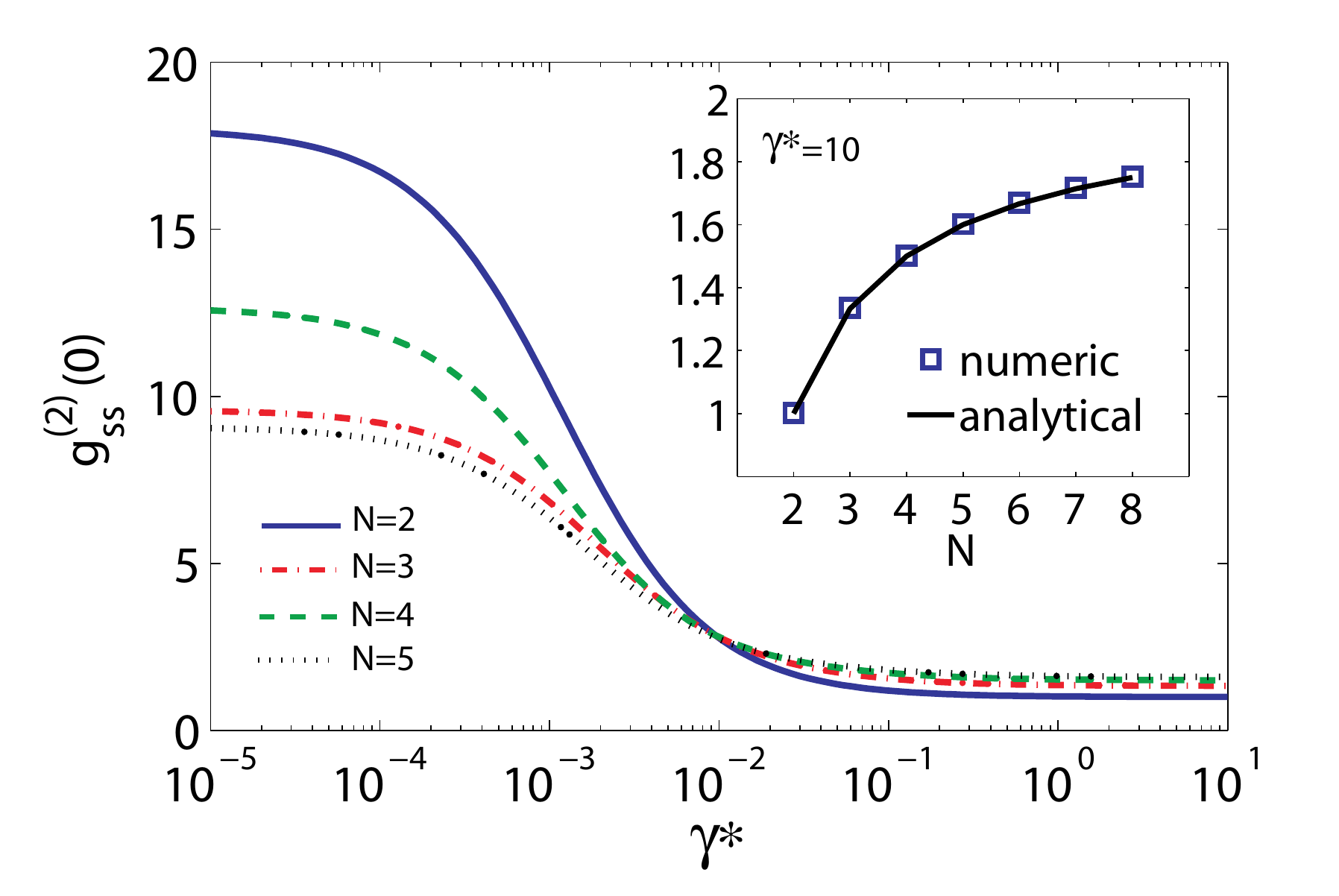}
\caption{ Zero time delay second-order correlation as a function of
pure dephasing rate, $\gamma^{\ast}$ (normalized to $g$), at fixed
$P_x=0.001g$ and $\kappa=100g$. The inset shows the behavior as a
function of $N$ at large $\gamma^{\ast}$. } \label{fig:deph_dep}
\end{center}
\end{figure}

We consider now the case where the cavity finesse is decreased down
to a point where the coupling of each emitter to the mode, $\Gamma$,
becomes of the same order of magnitude of the coupling to its
individual decay channel, $\gamma$. As we have seen in Figs.~\ref{fig2}b 
and ~\ref{fig3}, this leads to totally uncorrelated emission events from 
the atomic dipoles. 
In this regime, which we define as the  ``individualization,'' 
the parity-induced oscillations vanish and the bunching degree
converges to $g^{(2)}(0) = 2(1-1/N)$ (see Fig.~\ref{fig3}, inset on the right), 
a behavior that we demonstrate below. Note that the same value for the
parameter $g^{(2)}(0)$ has been computed in the strong pumping case
\cite{hollandPRA2}. As a matter of fact, a large pump power leads to
total population inversion, a state where the dipoles
necessarily become uncorrelated. Moreover, it is interesting to show
in the context of the present work that an alternative way of obtaining 
a source of uncorrelated dipoles is to increase the pure dephasing rate, 
$\gamma^*$. 
The latter effect is shown  in Fig.~\ref{fig:deph_dep}, where a full 
numerical solution based on direct integration has been employed 
for $\kappa/g=100$. 
It is evident that individualization is reached as soon
as $\gamma^*>\Gamma$, washing out the oscillations and leading to
the same limit for the second order correlation parameter, as clearly 
displayed in the inset.

For the sake of completeness, we have computed the steady state
second-order correlation  $g^{(2)}(\tau)$, for any delay $\tau$.
Adiabatic elimination is valid and this quantity can be calculated as
a function of the atomic operator, namely
$g^{(2)}(\tau)={\langle J_+ (0) J_+ (\tau) J_- (\tau) J_-(0)\rangle}/{\langle
J_+ (0)J_-(0) \rangle^2}$. We make the assumption that the steady
state of the ensemble can be factorized in the uncoupled basis, which
is reasonable as each emitter behaves individually. Thus, developing
(see App.~\ref{B}) the expression of $J_\pm$ with respect to the
individual atomic operator $\sigma_\pm^i$ leads to
\begin{equation}
g^{(2)}(\tau) = \frac{\sum_i g_i^{(2)}(\tau) I_i^2 + \sum_{i \neq k}  I_i I_k \{1+g_i^{(1)}(\tau)^* g_i^{(1)}(\tau)\}
}{(\sum_i I_i)^2 }\label{eq:cct} \, ,
\end{equation}
where we have introduced the intensity per emitter, the
first and the second order correlation functions for the quantum emitter $i$, as
$I_i =  \langle \sigma_+^i(0) \sigma_-^i (0) \rangle$,
$g_i^{(1)}(\tau)=\langle \sigma_+^i(\tau)\sigma_-^i(0)\rangle /
I_i$, $g_i^{(2)}(\tau)=\langle
\sigma_+^i(0)\sigma_+^i(\tau)\sigma_-^i(\tau)\sigma_-^i(0)\rangle
/ I^2 _i$, respectively. 
As initially mentioned in seminal lectures \cite{cct} and
papers \cite{Loudon}, and as it clearly appears in
Eq.~(\ref{eq:cct}), the second order correlation function involves
the emission of pairs of atomic excitations. 
When $\tau=0$, this expression simplifies to
\begin{equation}
g^{(2)}(0)=2(1-1/N) \, , \label{g2_right}
\end{equation}
confirming our numerical simulations (see right inset of
Fig.~\ref{fig2}). The bunching at small delays is due to an
interference resulting from the indistinguishability of the atoms in
the emitting pairs \cite{cct}. Note that surprisingly, this behavior was never observed
experimentally. As a matter of fact, the interference is washed out as
soon as individual first order coherence, $g_i^{(1)}(\tau)$,
vanishes~\cite{Jakeman77,Carmichael78,paper_g2,Paul82,Loudon}. This
phenomenon takes place on the timescale of dephasing processes,
leading to the following expression:
\begin{equation}
g^{(2)}(0)\simeq 1-1/N\label{truc}
\end{equation}
eventually corresponding to a particle-like
behavior. In fact, let's consider $N$ independent sources 
each emitting classical identical particles one-by-one. 
Such a model-source is characterized by a total emission intensity  
$I=\sum_i I_i$  with $\langle
I_i(t) I_j(t+\tau)\rangle=0$ if $i\neq j$. In the stationary regime
we have $\langle I_i(t)\rangle=\langle I_i(t+\tau)\rangle=I_0$, since
we suppose that the $N$ sources are characterized by the same main current.
The total second order correlation signal is therefore
\begin{equation}
g_{\textrm{tot}}^{(2)}(\tau)=\frac{\langle I(t)I(t+\tau)\rangle}{\langle I(t)\rangle \langle I(t+\tau)\rangle}
\end{equation}
which becomes
\begin{equation}
g_{\textrm{tot}}^{(2)}(\tau)=\frac{\sum_{i}g_i^{(2)}(\tau)+N(N-1)}{N^2}.
\end{equation}
However, our $N$ idealized sources are single particle emitters i.e.
$g_i^{(2)}(0)=0$. This clearly mimics the quantum-like behavior of
two-level systems. Therefore, we get Eq.~(\ref{truc}).
Interestingly, from basic textbook quantum optics we obtain for  a
pure $|n\rangle$ photon Fock state the expression  
$g^{(2)}(0)=1-{1}/{n}$.
The result is therefore also intuitive from this point of view.

The limit given in Eq.~(\ref{truc}) was already observed in the 
pioneering antibunching experiment by Kimble, Dagenais and 
Mandel with flying sodium atoms coherently driven~\cite{paper_g2},  
and it was theoretically interpreted in the following years
\cite{Jakeman77,Carmichael78,paper_g2,Paul82,Loudon}. In these
references, the source of dephasing was mainly attributed to the
finite size of the atomic source with a random number (i.e. a Poisson
distribution) of emitters inducing a spatial averaging out beyond
the typical coherence volume of the fluorescence light
\cite{Carmichael78,paper_g2}. 
In nowadays experiments with solid state emitters confined in 
nanometre-size systems, this problem could be ultimately overcome 
by considering a good control of the optical detection. 
However, notice that dephasing processes are always
present in solid state quantum emitters (due for example to
coupling with phonons in solid matrices), and that the measurement
of $g^{(2)}(0)$ involves actually an average over finite values of
$\tau$ near $\tau\simeq 0$. If the typical dephasing time
$\tau^*=1/\gamma^*$ is larger than the time resolution of the HBT
setup (e.g. the jitter time of single photon detectors) this may
explain why experiments reported so far with different quantum
emitters~\cite{Brouri2000,Kitson98,Hui2009,Weston2002,Cuche2010,Wrachtrup2010,Gerardot2005}
tend to evidence a behavior that is closer to Eq.~(\ref{truc})
rather than the present prediction described by Eq.~(\ref{g2_right}). Observing 
the $2(1-1/N)$ behavior,  and its transition to $1-1/N$, could 
be a challenging issue, e.g. requiring a control over the local temperature 
of the emitters.

\section{Physical realizations}

The results presented in this work suggest the intriguing 
possibility of studying the effects of cooperative emission with 
a mesoscopic number of quantum emitters in a high-finesse resonator.  
Moreover, depending on the specific physical implementation of 
the proposed model, different regimes of cavity QED might be explored and 
characterized in terms of the statistical properties of the emitted radiation. 
To this purpose,  we stress the relevance of Fig.~\ref{fig3} in providing a roadmap to 
experimental investigations. In fact, depending on the specific system of few
quantum emitters coupled to a single cavity mode, different regimes of $\kappa/g$
might be realized.

However, some of the requirements that we assumed in our analysis might
be difficult to be simultaneously realized in a physical implementation of the 
proposed model. Namely, we assumed identical quantum emitters,
equal dissipation rates, low pumping of each individual emitter as compared to 
its coupling to the cavity mode.  In order to realize the conditions
$\gamma^*, \gamma <P_x < \Gamma$, each quantum emitter should 
have a high beta factor (quantifying the coupling to the cavity loss channel).
Usually, this is quite challenging with solid state artificial atoms, like
quantum dots, where inhomogeneous broadening (which can be related 
to $\gamma^*$) would inevitably mix the different $J$ states. Atomic systems
in optical resonators would certainly offer an ideal implementation for what
concerns the atomic-like properties, but it would be difficult to have a mesoscopic
number of such atoms at fixed spatial positions.
Alternatively, atomic-like impurities in solid state resonators, e.g. semiconductors,
might represent a valid scenario for this type of physics to be investigated. 

As a first example, we consider semiconductor quantum dots,
especially made in III-V materials, as a promising 
approach to coupling in a deterministic way~\cite{hennessy07,dousse2008} a 
few quantum emitters to the same cavity mode, e.g. using
photonic crystal cavities~\cite{strauf06}.
Such systems can have $g\simeq 0.1$ meV \cite{hennessy07,faraon08} 
and $\kappa\simeq 0.01$ meV~\cite{derossi08apl}. 
We can thus expect to be in the range 
$\kappa / g \sim 0.1$ for such a physical implementation, 
where subradiant vs. lasing behaviors could be studied under weak driving conditions.
As these nanostructures are usually grown randomly, and strong inhomogeneous
fluctuations between different quantum emitters occur, techniques
have been recently developed to post-process the sample and obtain
more homogeneous ensembles~\cite{rastelli}. 
Electrical control of different quantum emitters has also been shown
to provide an effective tuning mechanism~\cite{arne}.
Moreover, growth of pyramidal 
dots in pre-patterned positions~\cite{pelucchi} is a very promising technique to
place more than one such nanostructures in the antinodes of
the cavity field in order to maximize coupling.

For active centers in solid-state resonators, the coupling to
the cavity mode is usually several orders of magnitude smaller, dictated by the smaller
oscillator strength of an atomic-like impurity as compared to a quantum dot. In
such case, the conditions for high-$\kappa /g $ values might be investigated, i.e.
the crossover from subradiant to superradiant emission, and maybe to individualization.
Recently, single to few Nitrogen-Vacancy centers in diamond have been implanted 
in pre-determined spatial positions~\cite{wrachtrup2009}. Resonators in diamond have 
already been realized with quality factors ranging from few hundreds in early 
attempts~\cite{wang07apl1,wang07apl2} up to 4000 more recently~\cite{faraon2011}. Furthermore,
NV-centers are not the only type of active defects in diamond, 
see e.g. recent advances with Cr-centers \cite{prawer2011}
and Silicon-vacancies (SiV centers)~\cite{weinfurter2006}.
Among solid state emitters, we also include single molecules in standing wave 
resonators made from a Bragg mirror and a resolution limited focused beam 
through an optical fibre~\cite{toninelli2010apl}. The latter approach has the 
advantage that the cavity can be formed at any arbitrary position, once the
single molecules have been selected on the Bragg mirror surface. Such quantum
emitters display radiation-limited zero-phonon line, thus showing very limited
pure dephasing rate with almost radiation-limited linewidths~\cite{toninelli2010oe}. 
 
Finally, a promising platform to realize a controlled experiment might be with 
circuit QED devices, where the coupling of a few qubits to a single resonator
has already been realized~\cite{dicarlo2010nat}. High tunability of the qubits physical
properties, such as transition energy, dissipation rates, pure dephasing and effective
oscillator strength, make these systems particularly attractive to closely realize
the model in Eqs. (\ref{TC_H}-\ref{ME}), and to experimentally address the physical results 
reported in the present work.

\section{Conclusions and perspectives}

We have presented an extensive theoretical analysis of the quantum 
statistical properties of a source of electromagnetic radiation made
of few two-level emitters coupled a single resonator mode. We have
assumed a model in which the two-level systems are incoherently pumped
by a weak external drive, and dissipate their energy through spontaneous
emission in free space, pure dephasing, or other loss channels that
can be described within a Liouvillian formalism. In this sense, the present work
should be considered as an extension of our previous report on single
emitter cavity QED~\cite{ours}, where we have analyzed the crossover
from a single-photon source (emitting anti-bunched radiation) to a 
single-emitter lasing regime. 

However, the case of more than one emitter studied here presents 
a much richer phenomenology depending on the specific light-matter coupling 
regime of each single emitter with the common radiation field, 
which can be identified from the statistical properties of the emitted 
radiation. Throughout this work, we have assumed a tuning parameter given 
by the ratio of the cavity decay rate to the single 
emitter-cavity coupling strength,  which we could numerically investigate 
by spanning over 10 decades. 
As a function of this parameter, we have clarified the crossover from lasing 
(low cavity decay rate as compared to emitters cavity coupling) to 
chaotic light emission (very large cavity decay rate), going
through sub- and super-radiance in steady state. 
Although in the superradiant regime the emitted radiation would present 
a close-to-Poissonian statistics, we have explicitly shown that it would 
not reach the level of coherence displayed by a laser source. 
Between these two regimes, sub-radiant emission takes place and 
the emitted radiation undergoes super-Poissonian intensity fluctuations. 

We have shown that the strong bunching in the degree of the second-order 
coherence oscillates with the parity of the number of emitters, eventually 
converging to a single value in the large number limit. 
We have clarified the origin of this novel behavior by using a suitable basis
for the system of few emitters, showing that it is due to the presence of
extra-loops in the radiative cascade from subradiant states when the number
of emitters is odd, thus reducing the degree of bunching with respect the
even case. 
Finally, we have also studied the conditions for reaching an incoherent 
source of chaotic radiation, which can be done either by coupling the 
ensemble of emitters to a very leaky cavity, or by enhancing the 
pure dephasing rate (a feature that is quite natural for solid-state 
quantum emitters).

As a last remark, notice that the cavity finesse has been the only parameter
to be significantly changed in order to produce very different output fields. 
Recently, fast dynamical control of this parameter has been demonstrated 
in different scenarios~\cite{Switch1,Switch2}, suggesting that a system 
composed of few emitters inside a resonant cavity could be a versatile source 
of very different quantum states of light.
An immediate and important extension of this work related to its experimental 
and technological implications would be to investigate the effects caused by the 
inhomogeneous distribution of the quantum emitters, which is a typical setting 
for smiconductor quantum dots. A further seemingly important extension would 
be to explore the consequences of the ultra-strong coupling regime (available 
nowadays for superconducting qubits in microwave resonators) on the statistics 
of the emitted radiation. In the latter scenario, anti-rotating wave terms in
the hamiltonian (\ref{TC_H}) must be taken into account~\cite{deliberato} in order 
to properly get the emitted radiation and statistics.

\begin{acknowledgments} 
A.A., D.G., and S.P. acknowledge financial support
from project NanoSci-ERANET project ``LECSIN''. A.A. acknowledges the
support of the French ANR through the project "CAFE". M.F.S. would
like to thank the National Research Foundation and the Ministry of
Education of Singapore. This work is part of the National Institute
of Science and Technology of CNPq, Brazil, and it has been partly
carried out at the Centre for Quantum Technologies in Singapore,
which is gratefully acknowledged for the kind hospitality. The
authors acknowledge M. Richard for stimulating discussions, and G.
Nogues for his precious help with QMC simulations.
\end{acknowledgments}

\begin{figure}[t]
\setlength{\unitlength}{8pt}
\begin{picture}(30,18)(1,1)
\linethickness{0.5pt}
\put(3,17){\line(1,0){1}}
\put(3,16){\line(1,0){1}}
\put(3,16){\line(0,1){1}}
\put(4,16){\line(0,1){1}}
\put(3,16){\makebox(1,1){1}}
\put(4.6,16.3){$\otimes$}
\put(6,17){\line(1,0){1}}
\put(6,16){\line(1,0){1}}
\put(6,16){\line(0,1){1}}
\put(7,16){\line(0,1){1}}
\put(6,16){\makebox(1,1){2}}
\put(7.6,16.3){$\otimes$}
\put(9,17){\line(1,0){1}}
\put(9,16){\line(1,0){1}}
\put(9,16){\line(0,1){1}}
\put(10,16){\line(0,1){1}}
\put(9,16){\makebox(1,1){3}}
\put(11.6,16.3){$=$}
\put(14,17){\line(1,0){1}}
\put(14,16){\line(1,0){1}}
\put(14,16){\line(1,0){1}}
\put(14,14){\line(1,0){1}}
\put(14,14){\line(0,1){3}}
\put(15,14){\line(0,1){3}}
\put(14,16){\makebox(1,1){1}}
\put(14,15){\makebox(1,1){2}}
\put(14,14){\makebox(1,1){3}}
\put(12,11){$\oplus$}
\put(14,12){\line(1,0){2}}
\put(14,11){\line(1,0){2}}
\put(14,10){\line(1,0){1}}
\put(14,10){\line(0,1){2}}
\put(15,10){\line(0,1){2}}
\put(16,11){\line(0,1){1}}
\put(14,11){\makebox(1,1){1}}
\put(15,11){\makebox(1,1){2}}
\put(14,10){\makebox(1,1){3}}
\put(26,11){$i\!=\!1$}
\put(19,11){$\vec{\lambda}\!=\!(2,1)$}
\put(6,11){$\big(J\!=\!\frac{1}{2}\big)$}
\put(12,7){$\oplus$}
\put(14,8){\line(1,0){2}}
\put(14,7){\line(1,0){2}}
\put(14,6){\line(1,0){1}}
\put(14,6){\line(0,1){2}}
\put(15,6){\line(0,1){2}}
\put(16,7){\line(0,1){1}}
\put(14,7){\makebox(1,1){1}}
\put(14,6){\makebox(1,1){2}}
\put(15,7){\makebox(1,1){3}}
\put(26,7){$i\!=\!2$}
\put(19,7){$\vec{\lambda}\!=\!(2,1)$}
\put(6,7){$\big(J\!=\!\frac{1}{2}\big)$}
\put(12,3){$\oplus$}
\put(14,4){\line(1,0){3}}
\put(14,3){\line(1,0){3}}
\put(14,3){\line(0,1){1}}
\put(15,3){\line(0,1){1}}
\put(16,3){\line(0,1){1}}
\put(17,3){\line(0,1){1}}
\put(14,3){\makebox(1,1){1}}
\put(15,3){\makebox(1,1){2}}
\put(16,3){\makebox(1,1){3}}
\put(19,3){$\vec{\lambda}\!=\!(3,0)$}
\put(6,3){$\big(J\!=\!\frac{3}{2}\big)$}
\put(14.5,15.5){\makebox(0,0){\line(1,2){1.8}}}
\put(14.5,15.5){\makebox(0,0){\line(1,-2){1.8}}}
\end{picture}
\caption{Total spin representations of three $\sigma=\frac{1}{2}$ spins
  with Young tableaux.}
\label{fig:youngdiagram}\label{fig6}
\end{figure}

\appendix

\section{Basis for the $N=3$ case}\label{A}

We hereby derive the basis states employed for the analytical
estimation of $g^{(2)}(0)$ in the three emitters case.
States with $J<J_{max}$ are degenerate as soon as $N>2$. 
We lift the degeneracy by exploiting the permutation symmetry in 
the free Hamiltonian, Eq. (\ref{TC_H}).  
Namely, we use symmetry-adapted state vectors related to the simultaneous 
diagonalization of the angular momentum $J$ (with $J_{\text{max}} = N/2$) 
algebra with the permutation group $P_N$ of N atoms \cite{Arecchi_PRA72}. 
We write the states in the coupled basis as 
\begin{equation}
\left| \begin{array}{cc}
J & \vec{\lambda} \\
M_J & i \end{array} \right\rangle \, ,
\end{equation} 
where $J$ and $M_J$ represent the collective dipole and its $z-$projection 
labels whereas $i$ enumerates the states belonging to  the $\vec{\lambda}$ 
permutation irreducible representation.
The derivation reduces to the use of standard branching rules for
Young tableaux with the theorem in Sec. V of Ref.~\cite{Arecchi_PRA72} 
or with the tensor method \cite{Wu-Ki Tung}.
Since we are dealing with a tensor product of two-dimensional 
Hilbert spaces, $\simeq \mathbb{C}^2$, then for 3 emitters the 
branching rules give three types of tableaux, out of which we 
must cancel those with more than two rows, as schematically
outlined in Fig.~\ref{fig6}. 
By means of Young symmetrizers, the coupled basis is easily
obtained in terms of the uncoupled states as

\begin{eqnarray}
&& \left|\frac{3}{2},\frac{3}{2}\right\rangle = \ket{e e e} \nonumber \\
&& \left|\frac{3}{2},\frac{1}{2}\right\rangle = \frac{1}{\sqrt{3}} \big( \ket{e e g} + \ket {e g e} + \ket{g e e} \big) \nonumber \\
&& \left|\frac{3}{2},-\frac{1}{2}\right\rangle = \frac{1}{\sqrt{3}} \big( \ket{e g g} + \ket {g g e} + \ket{g e g} \big) \nonumber \\
&& \left|\frac{3}{2},-\frac{3}{2}\right\rangle = \ket{g g g} \nonumber \\
&& \left| \begin{array}{cc}
\frac{1}{2} & \vec{\lambda}=(2,1) \\
\frac{1}{2} & i = 1 \end{array} \right\rangle = \frac{1}{\sqrt{6}} \big( 2 \ket{e e g} - \ket {e g e} - \ket{g e e} \big) \nonumber \\
&& \left| \begin{array}{cc}
\frac{1}{2} & \vec{\lambda}=(2,1) \\
-\frac{1}{2} & i = 1 \end{array} \right\rangle = \frac{1}{\sqrt{6}} \big( 2 \ket{g g e} - \ket {g e g} - \ket{e g g} \big) \nonumber \\
&& \left| \begin{array}{cc}
\frac{1}{2} & \vec{\lambda}=(2,1) \\
\frac{1}{2} & i = 2 \end{array} \right\rangle = \frac{1}{\sqrt{2}} \big( \ket {e g e} - \ket{g e e} \big) \nonumber \\
&& \left| \begin{array}{cc}
\frac{1}{2} & \vec{\lambda}=(2,1) \\
-\frac{1}{2} & i = 2 \end{array} \right\rangle = \frac{1}{\sqrt{2}} \big( \ket {g e g} - \ket{e g g} \big) \nonumber\end{eqnarray}

\section{Analytic solution for 3 emitters}\label{analytics}

In this appendix we show the analytical details for calculating 
the autocorrelation at  zero-time delay for the $N=3$ case.

The steady state solution to the system of equations relevant for the $g^{(2)}(0)$ 
calculation in the case of $3$ emitters gives:
\begin{eqnarray}
&& \rho_{3/2,-1/2} = \frac{P_x}{\Gamma} \rho_{3/2,-3/2} \nonumber \\
&& \rho_{3/2,1/2} = \frac{P_x (P_x + 2 \Gamma)}{4 \Gamma^2} \rho_{3/2,-3/2} \nonumber \\
&& \rho_{1/2,1/2} = \frac{P_x ( 2 P_x + 5 \Gamma)}{\Gamma (\Gamma + 6 P_x)} \rho_{3/2,-3/2} \\
&& \rho_{1/2,-1/2} = \frac{4 P_x + 3 \Gamma)}{\Gamma + 6 P_x} \rho_{3/2,-3/2} \nonumber \\
&& \rho_{3/2,3/2} = \frac{P_x ^2 ( 42 \Gamma^2 + 29 \Gamma P_x + 6 P_x ^2)}{12 \Gamma^3 (\Gamma + 6 P_x)} \rho_{3/2,-3/2} \nonumber\, .
\end{eqnarray}

Once we neglect the off-diagonal terms (see text in Sec.~\ref{3emitters}), the two 
degenerate states $\ket{1/2,\pm1/2;1}$, $\ket{1/2,\pm1/2;2}$ give the same values, 
and here they are considered simply as $\ket{1/2,\pm1/2}$.
The lowest matrix element is calculated from the constrain on the trace of the 
density matrix. It reads
\begin{equation}
\rho_{3/2,-3/2} = \frac{12 \Gamma ^3 ( \Gamma + 6 P_x)}{84 \Gamma^4 
+ 306 \Gamma^3 P_x + 201 \Gamma^2 P_x ^2 + 47 \Gamma P_x ^3 + 6 P_x ^4}\, .
\end{equation}
In powers of $P_x / \Gamma$ we finally obtain
\begin{eqnarray}\label{g2N3a}
g^{(2)}(0) \simeq&& \frac{14}{75} \frac{\Gamma}{P_x} + 2.067 - 3.501 \frac{P_x}{\Gamma} + 9.18 \left( \frac{P_x}{\Gamma} \right)^2 \nonumber \\
&& - 23.4 \left( \frac{P_x}{\Gamma} \right)^3 + \mathcal{O}\left( \frac{P_x}{\Gamma}\right)^4\, .
\end{eqnarray}
The same argument could have been used for the two-emitters case, for which the solution reads
\begin{equation}\label{g2N2}
g^2(0) \simeq \frac{4}{9} \frac{\Gamma}{P_x} + \frac{13}{27} - \frac{2}{27} \frac{P_x}{\Gamma} + \mathcal{O}\left( \frac{P_x}{\Gamma}\right)^2\, .
\end{equation}

In the case we are considering, namely $\Gamma / P_x = 40 $, the error in neglecting all the 
terms but the first one in $\Gamma / P_x$ is of about 3\% for Eq.~(\ref{g2N2}), while for eq. 
(\ref{g2N3a}) it would have been $\sim 30$ \%. 
This formally explains why in the two-emitters case the linear approximation works so well.
Indeed, setting $\Gamma / P_x = 40 $ in Eq.~(\ref{g2N3a}) we have $g^{(2)}(0) \simeq 9.46$ 
as given in the text.

\section{Individualization and chaotic limit}\label{B}

Developing the expression for $J_\pm$ with respect to individual atomic operators leads to
\begin{equation}
g^{(2)}(\tau) = \frac{\sum_{i,j,k,l} \langle \sigma_+ ^i (0)
\sigma_+ ^j (\tau)  \sigma_- ^k (\tau)  \sigma_- ^l (0)
\rangle}{\left(\sum_{k,l} \langle \sigma_+ ^k (0)  \sigma_- ^l (0)
\rangle \right)^2 }=\frac{\mathcal{N}}{\mathcal{D}}\, . \label{indiv01}
\end{equation}
We are considering the case where the atomic dipoles are
independent and incoherently excited. It means that whatever $i,j$
we have
\begin{eqnarray}
\langle \xi_i \xi_j \rangle &=& \langle \xi_i \rangle \langle \xi_j \rangle \nonumber \\
\langle \xi_i \xi_j \xi_i \rangle &=& \langle \xi_i \xi_i \rangle \langle \xi_j \rangle \nonumber \\
\langle \xi_i \rangle &=& 0 \, , \nonumber
\end{eqnarray}
being $\xi_i$ either $\sigma_- ^i$ or $\sigma_+ ^i$. It is immediate
to see that whenever $i \neq j,k,l$ (and permutations) the four
operators expectation value is zero. We are left with $i=j=k=l$ and
the case where the indices are equal to each other in pairs. 
As a consequence
\begin{eqnarray}
\mathcal{N}=\sum_{k} \langle \sigma_+ ^k (0)  \sigma_+ ^k (\tau)  \sigma_- ^k (\tau)  \sigma_- ^k (0) \rangle  \nonumber \\
&& \hspace{-4.0cm} +\sum_{k \neq l} \bigg( \langle \sigma_+ ^k (0)  \sigma_+ ^k (0) \rangle \langle  \sigma_- ^l (\tau)  \sigma_- ^l (\tau) \rangle \nonumber \\
&& \hspace{-3.0cm} +\langle \sigma_+^k (0)  \sigma_+^k (\tau) \rangle \langle  \sigma_-^l (\tau)  \sigma_-^l (0)\rangle \bigg)  \nonumber \\
&& \hspace{-5.0cm} = \sum_k g_k^{(2)}(\tau) I_k^2 + \sum_{k \neq l}  I_k I_l \big(1+g_k^{(1)}(\tau)^* g_l^{(1)}(\tau) \big)\, , \nonumber
\end{eqnarray}
where $I_k =  \langle \sigma_+^k(0) \sigma_-^k (0) \rangle$
For zero delay, the denominator $\mathcal{D}$ involves
$ \langle J_+ J_- \rangle =\sum_{i,j} \langle \sigma_+ ^i \sigma_- ^j \rangle $,
which simply writes $\sum_{i} \langle \sigma_+ ^i \sigma_- ^i\rangle = N I $,
as the dipoles are independent. As for the numerator, the case $ i=j=k=l $
gives no contribution as the field emitted by a single atom is antibunched,
the pairs $(i=k,j=l)$ and $(i=l,j=k)$ contribute with $ 2 N (N-1) $ terms
equal to $I^2$ (by definition, we have $g_l^{(1)}(0)=1$). Eventually,
one finds the $g^{(2)}(0) = 2(1-1/N) $ result mentioned in the text.
It is worth noticing that in the large number limit we have $g^{(2)}(0) = 2$
corresponding to a chaotic chaotic source \cite{cct,Loudon}.\\
It is also interesting to remark that with a distribution of $N$
classical electromagnetic emitters we still have
\begin{eqnarray}
g^{(2)}(\tau)=\frac{\sum_{k}g_k^{(2)}(\tau) +\sum_{k\neq l}g_k^{(1)}(\tau)(g_l^{(1)}(\tau))^\ast}{N^2}\nonumber\\
+1-\frac{1}{N}
\end{eqnarray}
However, the classical Schwartz inequality imposes $g_i^{(2)}(0)\geq
1$ and therefore we now obtain  for the $N$ emitters
\begin{equation}
g^{(2)}(0)\geq \frac{1}{N}+2(1-\frac{1}{N})=2-\frac{1}{N} \, .
\end{equation}

The value $2(1-1/N)$ is therefore somewhere intermediate between the
pure particle like behavior, $1-1/N$ (see main text), and the pure
classical undulatory prediction, $2-1/N$ (in agreement with the usual
Einstein interpretation of the fluctuation formula for the
black-body radiation).

\end{document}